\documentclass[preprint,prx,preprintnumbers,amsmath,amssymb,longbibliography]{revtex4-2}

\usepackage{times}
\usepackage{graphicx}
\usepackage{amsmath}
\usepackage{mathtools}

\usepackage[usenames,dvipsnames]{color}
\usepackage[normalem]{ulem}
\usepackage{comment}
\newcommand{\blue}[1]{{\color{Blue} #1}}

\newcommand{\oliver}[1]{{\color{black} #1}}
\newcommand{\oli}[1]{{\color{black} #1}}
\newcommand{\oliv}[1]{{\color{black} #1}}
\newcommand{\rader}[1]{{\color{black} #1}}

\begin{document} 

	\def\blue{\textcolor{black}}
	\def\gr{\textcolor[rgb]{0,0.6,0}}
	\def\br{\textcolor[rgb]{0.7,0.5,0}}
	\def\prp{\textcolor{black}}

	\def\StructGrapheneAbs{$7\sqrt{3}\times7\sqrt{3}$}
	\def\LDP{LD-phase}
	\def\HDP{HD-phase}
	\def\AB{{\it A-B}}
	\def\Ef{$E_{\rm F}$}
	\def\Tc{$T_{\rm C}$}
	\def\kpara{{\bf k}$_\parallel$}
	\def\kparax{{\bf k}$_{\parallel,x}$}
	\def\kparay{{\bf k}$_{\parallel,y}$}
	\def\kz{{\bf k}$_\perp$}
	\def\kperp{{\bf k}$_\perp$}
	\def\Gbar{$\overline{\Gamma}$}
	\def\Mbar{$\overline{\rm M}$}
	\def\Xbar{$\overline{\rm X}$}
	\def\dirGX{$\overline{\rm \Gamma}-\overline{\rm X}$}
	\def\dirMXM{${\rm M}{\rm X}{\rm M}$}
	\def\dirRMR{${\rm R}{\rm M}{\rm R}$}
	\def\dirXRb{${\rm X}{\rm R}$}
	\def\dirGMb{${\Gamma\rm M}$}
	\def\dirGY{$\overline{\rm \Gamma}\overline{\rm Y}$}
	\def\dirGK{$\overline{\rm \Gamma}\overline{\rm K}$}
	\def\dirGM{$\overline{\rm \Gamma}\overline{\rm M}$}
	\def\dirGS{$\overline{\rm \Gamma}\overline{\rm S}$}
	\def\dirGN{$\overline{\rm \Gamma}\overline{\rm N}$}
	\def\dirGMtic{$\overline{\rm \Gamma}_{\rm TiC}-\overline{\rm M}_{\rm TiC}$}
	\def\dirGKtic{$\overline{\rm \Gamma}_{\rm TiC}-\overline{\rm K}_{\rm TiC}$}
	\def\dirMKtic{$\overline{\rm M}_{\rm TiC}-\overline{\rm K}_{\rm TiC}$}
	\def\dirGMgr{$\overline{\rm \Gamma}_{\rm Gr}-\overline{\rm M}_{\rm Gr}$}
	\def\dirGKgr{$\overline{\rm \Gamma}_{\rm Gr}-\overline{\rm K}_{\rm Gr}$}
	\def\dirMKgr{$\overline{\rm M}_{\rm Gr}-\overline{\rm K}_{\rm Gr}$}
	\def\dirGMsc{$\overline{\rm \Gamma}_{\rm SC}-\overline{\rm M}_{\rm SC}$}
	\def\dirGKsc{$\overline{\rm \Gamma}_{\rm SC}-\overline{\rm K}_{\rm SC}$}
	\def\dirMKsc{$\overline{\rm M}_{\rm SC}-\overline{\rm K}_{\rm SC}$}
	\def\dirGNhalf{$\frac{1}{2}(\overline{\rm \Gamma}-\overline{\rm N})$}
	\def\pntG{$\overline{\rm \Gamma}$}
	\def\pntM{$\overline{\rm M}$}
	\def\pntK{$\overline{\rm K}$}
	\def\pntN{$\overline{\rm N}$}
	\def\pntGsc{$\overline{\rm \Gamma}_{\rm SC}$}
	\def\pntKsc{$\overline{\rm K}_{SC}$}
	\def\pntMsc{$\overline{\rm M}_{SC}$}
	\def\pntGtic{$\overline{\rm \Gamma}_{TiC}$}
	\def\pntKtic{$\overline{\rm K}_{\rm TiC}$}
	\def\pntMtic{$\overline{\rm M}\def\dirGX{$\overline{\rm \Gamma}-\overline{\rm X}$}{\rm TiC}$}
	\def\pntGgr{$\overline{\rm \Gamma}_{\rm Gr}$}
	\def\pntKgr{$\overline{\rm K}_{\rm Gr}$}
	\def\pntMgr{$\overline{\rm M}_{\rm Gr}$}
	\def\pntNhalf{ $\overline{\rm N}/2$ }
	\def\invA{\AA$^{-1}$}
	\def\DCgamma{${\rm DC}_{\overline{\Gamma}}$}
	\def\DCNhalf{${\rm DC}_{\overline{\rm N}/{\rm 2}}$}
	\def\root33{$\sqrt{3}\times\sqrt{3}$ {\it R}30$^\circ$}
	\def\RT3{$\sqrt{3}$}
	\def\aR{\alpha_{\rm R}} 
	\def\Ga{$\Gamma$}
	\def\GM{$\Gamma$M}
	
	\def\twobytwo{$2\times2$}
	\def\twotimestwo{$2\times2$}
	
	\def\CPB{CsPbBr$_3$}
	\def\CsPbI3{CsPbI$_3$}
	\def\CsPbBr3{CsPbBr$_3$}
	\def\CsPbCl3{CsPbCl$_3$}
	\def\MAPbI3{MAPbI$_3$}
	\def\MAPbBr3{MAPbBr$_3$}
	\def\MAPbCl3{MAPbCl$_3$}
	\def\MAPbX3{MAPb$X_3$}
	\def\CsPbX3{CsPb$X_3$}
	\def\MAPI{MAPbI$_3$}
	\def\MPB{MAPbBr$_3$}
	\def\CPB{CsPbBr$_3$}
	
	\def\Mp{M$^\prime$}

\title{\oliver{Large Jahn-Teller shifts and splittings observed in halide perovskite CsPbBr$_3$}}

\author{Maryam Sajedi,\blue{$^{1,\S}$} Maxim Krivenkov,$^{1,\S}$ Dmitry Marchenko,$^{1,\S}$ Saleem Ayaz Khan,$^{2}$ Andrei Varykhalov,$^{1}$ Jaime S\'anchez-Barriga,$^{1,3}$ Daniel M. T\"obbens,$^{1}$ Thomas Unold,$^{1}$ J\'an Min\'ar,$^{2}$ and Oliver Rader$^{1,\ast}$}

\affiliation{$^1$ Helmholtz-Zentrum Berlin f\"ur Materialien und Energie, Albert-Einstein-Str. 15, 12489 Berlin, Germany}
\affiliation{$^2$ New Technologies Research Centre, University of West Bohemia, 301 00 Pilsen, Czech Republic}
\affiliation{$^3$ IMDEA Nanociencia, c/ Faraday 9, Campus de Cantoblanco, 28049 Madrid, Spain}
\affiliation{$^\S$\ These authors contributed equally.}
\affiliation{$^\ast$\ Corresponding author.}

\begin{abstract}
{\oliver{\bf 
Despite the fundamental role of Jahn-Teller effects of first and second order in shaping structural and electronic properties, there is hardly any observation in angle-resolved photoemission of solids. 
\rader{In oxide and halide perovskites, band structure replicas have occasionally been reported as fingerprints of Jahn-Teller effects, but no accompanying energy shifts or splittings that would allow a conclusion about their origin.}
In CsPbBr$_3$, we uncover \rader{both} key signatures:  Upon cooling, orthorhombic replica bands emerge \rader{which had eluded earlier studies. This includes} an extra valence band maximum at  $\Gamma$ \rader{which can even be distinguished at room temperature. Most importantly,} band narrowing along $\Gamma$-$M$, a splitting at $\Gamma$, and the lifting of degeneracy between nonequivalent M points, all of several 100 meV, become apparent. Temperature-dependent x-ray diffraction, used as input for band structure calculations, links these effects directly to  tilts and rotations of the PbBr$_6$ octahedra. \rader{Our results uncover a strong electron-lattice interaction which is at the heart of so-far unresolved questions concerning polaronic transport, dynamic disorder, and exciton trapping in halide perovskites.}
  }} 
\end{abstract}

\maketitle

{\oliver{The Jahn-Teller (JT) effect is a fundamental phenomenon in physics and chemistry of a reduction of  structural symmetry to lift electronic degeneracy and minimize the energy of the system. In solids, the JT effect can drive structural phase transitions, lowering the  symmetry of the crystal  \cite{Kugel82,Imada98,Varignon19}. It practically extends the Peierls transition — originally described in one-dimensional systems — to higher dimensions and its itself extended by the second-order or pseudo-Jahn-Teller effect, which arises from near-degenerate electronic states \cite{Bersuker13}. Both mechanisms couple electronic states to lattice coordinates and  predict selective energy shifts and splittings in the electronic band dispersion. 
 
In transition-metal perovskites of $ABX_3$ structure, JT distortions can significantly alter exchange interactions, profoundly influencing magnetic properties and leading to the emergence of numerous magnetic and structural phases \cite{Kugel82}. These distortions encompass deformations of the $BX_6$ octahedra, as well as their tilts and rotations. Manganites, celebrated for their colossal magnetoresistance and metal-insulator transitions, are a prominent example of such materials \cite{Imada98}. 
Despite extensive study, no universal model for JT distortions has been established, especially as exchange interactions, electron-phonon coupling, and strong electron correlation often coexist in these materials \cite{Varignon19}.  
\oli{To differentiate observable fingerprints of the JT effect in solids, \rader{such as replica bands,} from its underlying mechanisms \cite{Varignon19}, it will be highly beneficial to identify JT distortions through band structure measurements. }
 
Angle-resolved photoemission spectroscopy (ARPES) is the key technique for probing the band structure of solids and has been extensively applied to perovskite oxides.
However, direct observations of JT effects in ARPES remain rare, with most studies identifying only indirect fingerprints. In colossal magnetoresistive manganite, replicas of the three-dimensional Fermi surface in ARPES have been attributed to octahedral tilts which define the rhombohedral phase \cite{Lev15}, analogous to observations in two-dimensional cuprate superconductors \cite{Man06} \rader{and ferromagnetic ruthenate \cite{Sohn21}}.  
 }}

{\oliver{Over the past decade, halide perovskite semiconductors have attracted significant interest for their high photovoltaic efficiency, long carrier lifetimes, defect tolerance, and tunable band gap, making them promising for photovoltaics, lighting, and radiation detection \cite{Ahmad22,Park22,Zhu24}. Despite differences, such as valence sums, they undergo the same orthorhombic–tetragonal–cubic phase transitions as oxides, driven by octahedral tilts and rotations \cite{LeeJH16, Young16}. \rader{In the temperature dependence, the orthorhombic phase appears at lowest temperature, followed by the tetragonal and the cubic phases} (see Supplementary Note 1). 

It has been suggested that the octahedral tilting and rotation   is crucial as it changes the orbital overlap at the band extrema which narrows both conduction and valence bands and increases the band gap \cite{Fil14, Young16, LeeJH16}.  Filip et el. \cite{Fil14} pointed out for MAPbI$_3$ (MA = methylammonium) that the $BX_6$ octahedra are rigid and almost perfect Platonic solids 
and that the band gap can be tuned in a large range from 1.3 to 1.8 eV (including the range of the Shockley-Queisser limit of 1.2 to 1.4 eV) by modifying only the bond angle through steric effects. This means that the ionic radii determine the octahedral tilts and rotations and hence the bond angle \cite{Fil14}. Indeed, Lee et al. \cite{LeeJH16} found a linear relationship between band gap and the tolerance factor, a criterion  based on treating all   ions as hard spheres  \cite{Kieslich15}. On the other hand, Linaburg et al. \cite{Lin17} noted that it is counterintuitive that the band gap of cubic MAPbCl$_3$ without tilts is larger than that of orthorhombic CsPbCl$_3$ with tilts.  Lee et al. \cite{LeeJH16}  suggested therefore that hydrogen bonding should be added as relevant factor.

Subsequently, Goesten and Hoffmann \cite{Goesten18} conducted a bonding analysis to relate the band gap to the chemical bonding. Two pairs of mirror bands of bonding and antibonding orbitals are identified which control the band gap. The valence band maximum (VBM) is nonbonding and located at the R point (corner point) of the Brillouin zone \cite{Goesten18}, confirming previous results \cite{Frost14}. Interestingly, Goesten and Hoffmann \cite{Goesten18} use only the cubic cell and therefore do not consider any tilt or rotation of octahedra. They   point out that an explanation for the orthorhombic distortion   of CsPbI$_3$ and CsPbBr$_3$ is lacking, and they conclude that there is no relation between band gap and geometry \cite{Goesten18}.
Given these contrasting perspectives, an investigation of the role of octahedral tilts and rotations for the electronic structure of halide perovskites is urgently needed. 

ARPES has so far confirmed the VBM at the R point and determined effective masses for MAPbBr$_3$  \cite{ZuJPCL19} and CsPbBr$_3$ \cite{Puppin20,Sajedi22}. We could clarify the absence of a Rashba effect at the VBM   \cite{Sajedi20} and of signs of large polarons   in the effective mass \cite{Sajedi22}. 
CsPbBr$_3$\ is orthorhombic at ambient temperature with \emph{Pnma} space group up to 361~K, tetragonal above, and cubic above 403~K \cite{hirotsu74}.
Nevertheless, only bands of the cubic structure have been observed in ARPES, with most data taken at room temperature and very few at low temperatures \cite{Puppin20,Sajedi22}. Orthorhombic   bands in ARPES have most recently   been observed in MAPbI$_3$\ \cite{Park24}, but their absence in CsPbBr$_3$ is unexpected and potentially important for the following reason: ARPES has a small probing depth \rader{of 5--10 \AA\ for photon energies of 10--100 eV}, and based on this, it has been concluded that at room temperature, the surface of MAPbI$_3$\ is cubic, while the bulk is tetragonal \cite{ZuJPCL19}. This conclusion appears to be in line with the photoluminescence of MAPbBr$_3$, which for 2~nm nanocrystals behaves differently from 4~nm ones: The smaller nanocrystals remain cubic at all temperatures due to the more decisive influence of the surface, whereas the larger ones undergo the expected bulk-like phase transitions \cite{LiuL19}. \rader{A recent ARPES study at 100 K reported the absence of orthorhombic bands in CsPbBr$_3$ and aimed to provide a fundamental understanding of this absence \cite{ChungNP24}.}

In the present article, we will at first resolve   the apparent disagreement between the structural determination by x-ray diffraction (XRD) and ARPES by demonstrating the orthorhombic bands. \rader{We demonstrate this at low temperature ($\ge12$ K), where the orthorhombic bands are prominent, and at room temperature, where they are much weaker.} Subsequently, we will reveal energy shifts and splittings as candidates for a second-order JT effect. We conduct temperature-dependent XRD as input for our DFT calculations and confirm the JT origin of the observed energy shifts and splittings. 
}}
 

{\it Methods.} CsPbBr$_3$\ single crystals have been homegrown by the antisolvent vapor-assisted crystallization  method \blue{(see Suppl.~Note~2}).
ARPES has been performed using hemispherical electron energy analyzers and synchrotron radiation at the ARPES-1$^2$ (Scienta R8000) and Spin-ARPES (Scienta~R4000) instruments at BESSY II using linearly polarized undulator radiation. The base pressure of the instruments was better than $2\times10^{-10}$ mbar and the spot size in ARPES of $\sim$100 $\mu$m. XRD has been performed as a function of temperature (see Suppl.~Note~5). The band structure calculations based on density functional theory were conducted with the Vienna ab initio Simulation Package (VASP) (see Suppl.~Note~3).

\begin{figure}[!ht]
	\centering
	\includegraphics[width=0.98\textwidth]{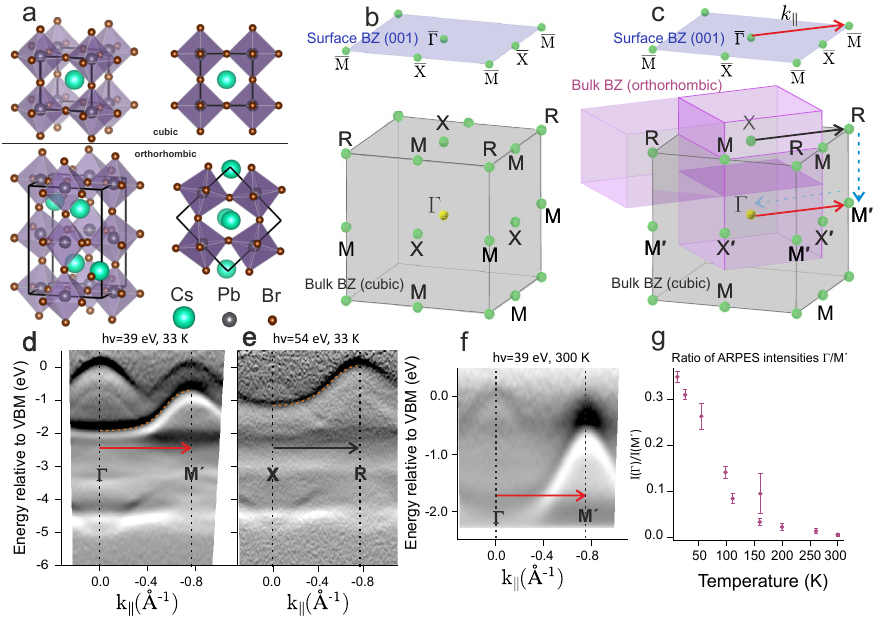}
	\caption{\label{orth} 
{\bf Orthorhombic bands at low temperature and room temperature.} 
{\bf a} Cubic and orthorhombic structures. The unit cell (black lines) increases 4 times due to tilts and rotations of the octahedra. 
{\bf b} Cubic bulk and corresponding surface Brillouin zone and {\bf c} orthorhombic Brillouin zone. 
Solid arrows mark the probed $k$-directions. 
{\bf d,e} ARPES along $\Gamma$\Mp\ and XR at low temperature (33~K). Replicas of the VBM appear at $\Gamma$ and \Mp\ backfolded by reciprocal lattice vectors (blue dashed lines) shown in {\bf c}. 
The primed “ $^\prime$ ” letters denote non-equivalent high-symmetry points. {\bf d-f} show the first derivative of ARPES intensities. 
{\bf f} The VBM replica at $\Gamma$\ can also be discerned at room temperature, where its intensity is very weak. 
{\bf g} ARPES intensity behavior of the VBM replica at $\Gamma$ normalized to the intensity of the local maximum at \Mp\ (saddle point) versus temperature.
}
\end{figure}


{\it Cubic versus orthorhombic bands.}
First, we demonstrate that strong zone folding effects at the lowest temperatures confirm the presence of the orthorhombic lattice. We then analyze the temperature dependence of the additional features.
Figure.~\ref{orth}~{\bf a} presents the cubic high-temperature and orthorhombic low-temperature structure of CsPbBr$_3$\ and halide perovskites in general. The bulk and surface Brillouin zones of the cubic structure are shown in Fig.~\ref{orth}~{\bf b}, while the orthorhombic Brillouin zone, which is four times smaller, is depicted in Fig.~\ref{orth}~{\bf c}. Measurements along bulk $\Gamma$\Mp\ and XR directions (red and yellow arrows in Figure~2~{\bf c-e})  were performed at 33~K, using photon energies of 39~eV and 54~eV, respectively. In Fig.~\ref{orth}~{\bf d}\blue{,} the $\Gamma$\Mp\ band is highlighted by orange dashed lines, while Fig.~\ref{orth}~{\bf e} marks the XR band in the same manner. All other bands are primarily replicas due to the orthorhombic lattice reconstruction\blue{,} which maps the XR direction onto the $\Gamma$\Mp\ direction and vice versa \blue{(see Suppl.~Note~4)}. 
The local maximum at \Mp\ is in fact a saddle point, as evident along RM (see Fig. S2). The VBM at the R point also becomes visible at the $\Gamma$ point\blue{;} one can regard it as a double zone folding, e.g., from R to \Mp\ and then to $\Gamma$, as indicated in Fig.~\ref{orth}~{\bf c} by blue dashed arrows. 
\rader{Our observations of orthorhombic replicas is in agreement with a} recent ARPES report for MAPbI$_3$ \rader{(which is orthorhombic below 165 K)} \cite{Park24} \rader{but contrasts recent ARPES and theory data for CsPbBr$_3$ which do not report orthorhombic replicas at 100 K and predict their absence in normal emission (${\bf k}_\parallel=0$) \cite{ChungNP24} (see Supplementary Note 4).}

\rader{Now we will turn to room temperature.} Until now, ARPES measurements at room temperature have only revealed bands associated with the cubic structure \cite{Puppin20,Sajedi22}, despite the orthorhombic lattice structure  determined by XRD \cite{hirotsu74,rodova03}. The absence of orthorhombic bands in ARPES at room temperature is unexpected, and we seek to resolve this inconsistency. To investigate this, we perform room-temperature ARPES measurements under the same conditions as in  Fig.~\ref{orth}~{\bf d}, where the replica band is relatively pronounced.  We then plot the derivative of the logarithm of the ARPES intensity with respect to energy, with the resulting data shown in Fig.~\ref{orth}~{\bf f}. While the replica at $\Gamma$ remains weak at room temperature (see the temperature-dependent intensity in Fig.~\ref{orth}~{\bf g}), it is still discernible.
{\oliver{
The appearance of the orthorhombic replica bands could serve as a fingerprint for a JT effect but as such it does not convey extra information about the temperature dependence of the electronic structure. Nevertheless, this observation reconciles ARPES with XRD findings and establishes ARPES as a viable tool for further investigations into the bulk electronic properties of CsPbBr$_3$ and a potential JT effect.
}}

  \begin{figure}[!ht]
  	\centering
  	\includegraphics[width=0.98\textwidth]{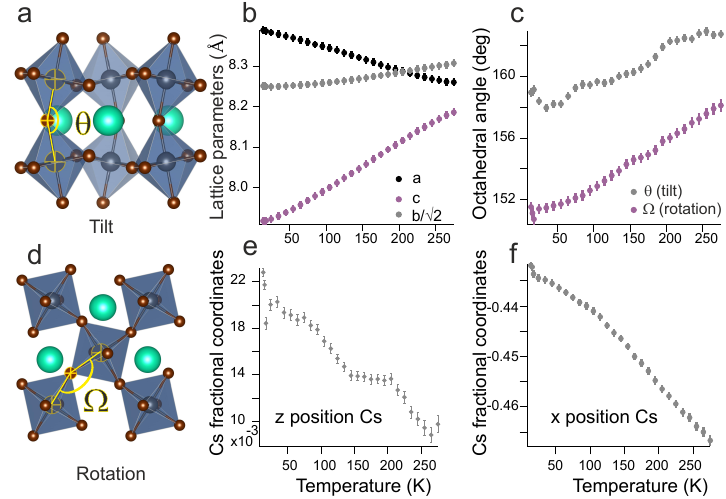}
  	\caption{\label{XRD} 
{\bf Temperature-dependent displacement of octahedra.}
{\bf a,d} Illustration of the octahedral tilt $\theta$ and rotation $\Omega$ angles. 
{\bf b} Temperature dependence of the unit-cell parameters from the XRD data, 
{\bf c} Pb-Br-Pb tilt and rotation angles, 
{\bf e,f} position of the Cs ion.}
  \end{figure}


{\it Energy shift and splitting.} We now compare the electronic structure of the low-temperature orthorhombic phase with that of the room-temperature phase, which we refer to as pseudocubic, given the weakness of the orthorhombic replica features and the presence of strong local thermal fluctuations \cite{Canelli22}. 

\begin{figure}[!ht]
	\centering
	\includegraphics[width=0.55\textwidth]{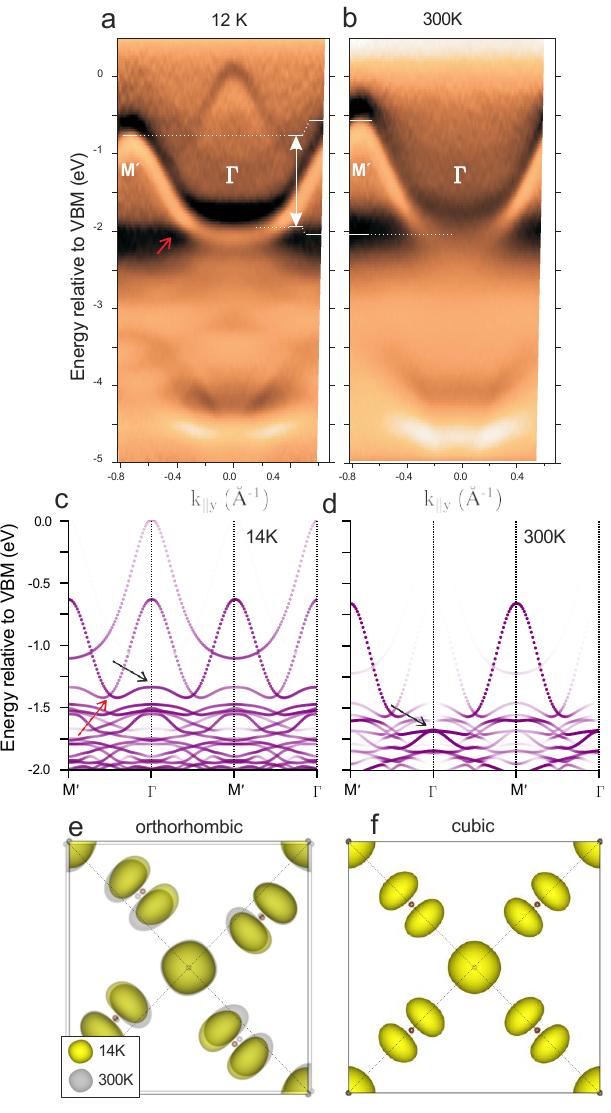} 
	\caption{\label{orbital} {\bf Temperature-dependent electronic band narrowing, and \oliv{splitting}.} 
		{\bf a,b} ARPES (first derivative) along $\Gamma$\Mp\ (i.e., along the red arrow in Fig.~\ref{orth}) showing a band narrowing by $\sim$300~meV \oliv{or 20\%} at low temperature, $h\nu=39$~eV.   
		{\bf c--d} DFT calculation along $\Gamma$\Mp\ based on the XRD data of Fig.~\ref{XRD}~{\bf b},{\bf c},{\bf e}, and {\bf f}. The calculated band narrowing along $\Gamma$\Mp\ confirms the experimental one in {\bf a,b}.
		{\bf e,f} Real-space orbitals at the \Mp\ saddle point for {\bf e} the 14~K and 300~K cases superimposed and {\bf f} the cubic case.}
\end{figure}

Figure.~\ref{orbital}~{\bf a,b} compares the \Mp$\Gamma$\Mp\ direction at 12~K and 300~K. In addition to the VBM replica at $\Gamma$, we observe a downward shift of the saddle point at \Mp\ by 200~meV upon cooling. (Note that the saddle point at \Mp\ [Fig. S2] appears  as a local maximum along \Mp$\Gamma$.) Furthermore, the band bottom shifts upward  \oliv{by $\sim$100 meV}, resulting in a splitting and a kink (red arrow). Since the band bottom is difficult to resolve at 300~K, we estimate that the overall band narrowing amounts to about 
300~meV \oliv{or 20\%}. This substantial value suggests a pronounced effect, pointing to significant structural rearrangement. We propose that the observed energy shift is a manifestation of the JT effect.
 
 To analyze this further, we conducted temperature-dependent XRD experiments and used the obtained structural data as input for DFT calculations.
Accurate structural parameters of bulk CsPbBr$_3$\ were derived from our  XRD measurements (see Fig.~\ref{XRD}~{\bf ~b,c,e,f}). The general trend of the octahedral tilt and rotation angles (defined in  Fig.~\ref{XRD}~{\bf a,d} and Suppl.~Note~1) shows an increase  (deviating further from 180$^{\circ}$) as the temperature decreases. This increasing distortion enhances the intensity and visibility of replica bands in ARPES. As mentioned above, the crystal structure of CsPbBr$_3$ has been identified as orthorhombic with space group  \textit{Pnma} at room temperature and below  \cite{hirotsu74,rodova03,ZhangRSC17}. Our powder diffraction data fully support this classification, and we adopt this model as the basis for our DFT calculations. The validity of this approach is unaffected by a recent single-crystal XRD study \cite{Liu21XRD}, which assigned a reduced subgroup symmetry and monoclinic space group  \textit{Pm} \blue{(see Suppl.~Note~5)}.

\indent Fig.~\ref{orbital}~{\bf c,d} presents the DFT band structures calculated using the atomic structures derived from our XRD data for the orthorhombic phase, unfolded to the cubic Brillouin zone. We used XRD structures  at 14~K and 300~K for these calculations. The replica of the VBM at $\Gamma$, predicted in Fig.~\ref{orbital}~{\bf c}, corresponds exactly to the experimental observation in Fig.~\ref{orbital}~{\bf a}. Additionally, the narrowing of the bandwidth observed experimentally is also predicted by the DFT calculations: from the saddle point at \Mp\ (a local maximum along \Mp$\Gamma$) to the kink (red arrow in the experimental data in Fig.~\ref{orbital}~{\bf a}, located halfway between \Mp\ and $\Gamma$), the bandwidth narrows by 170~meV in the calculation. When considering only the intense bands as the band bottom (indicated by the black arrows in Fig.~\ref{orbital}~{\bf c,d}), the calculated band narrowing increases to 320~meV. 
	
These findings suggest the following Jahn-Teller effects: (i) The octahedral tilt and rotation contribute to the reduction in bandwidth, which we interpret as second-order JT effect. (ii) The splitting near the band bottom at $\Gamma$ may indicate the lifting of degeneracy, but it is most likely caused by a change in orbital hybridization, thus a second-order JT effect as well. (iii) The pronounced kink observed is a sign of the zone folding resulting from the orthorhombic structure. 

Figure~\ref{orbital}~{\bf f} shows the calculated charge density at the \Mp\ saddle point for the cubic phase. We observe the Pb 6s and Br 4p orbitals in an $sp\sigma^*$ configuration, which is consistent with previous predictions \cite{Goesten18}. In the orthorhombic calculation (Fig.~\ref{orbital}~{\bf e}), the Br~4p orbitals are displaced at 300~K and even more so at 14~K, where they also undergo a slight rotation. This tilt is in agreement with prior calculations for MAPbI$_3$ \cite{LeeJH16}, although the effect of the tilt and rotation on the electronic structure was significantly overestimated, with predictions suggesting downward shifts at the M point of about 0.5~eV \cite{LeeJH16}.

\begin{figure}[!ht]
	\centering
	\includegraphics[width=0.98\textwidth]{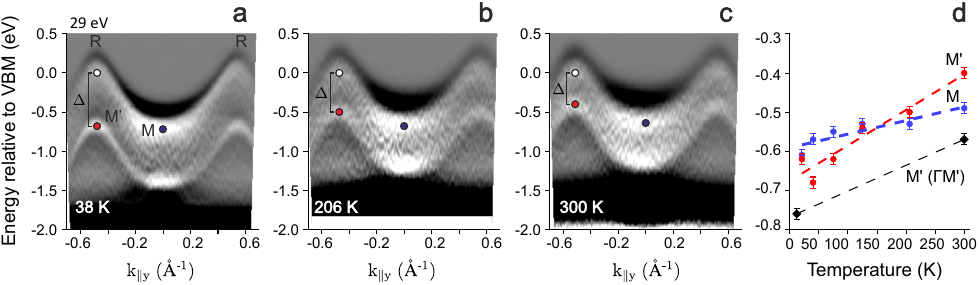} 
	\caption{\label{T_dep} {\bf Second-order Jahn-Teller effect on the M point.} {\bf a-c} Temperature-dependent ARPES data along RMR (second derivative). Two nonequivalent M points, M and \Mp\ are traced with temperature. The split-off band  below the VBM is due to an umklapp from \Mp. The M point is probed directly at $k_{\parallel y}=0$ and changes much less with temperature \oliv{than \Mp}. {\bf d} \prp{Energy positions of M and replica of \Mp\ points, values for M \oliv{have been} shifted up by 150 meV due to k$_z$ offset \oliv{(see text)}. Black points are extracted from Fig.~\ref{orbital}~{\bf a,b}.} }
\end{figure}

A remaining question concerns the degeneracy lifted by the JT effect. While we observe a splitting at $\Gamma$, no such splitting is seen at M. This is because there are two or three nonequivalent M points in the orthorhombic lattice, and in ARPES we probe one at a time. If we can demonstrate that two of these points are non-degenerate at low temperature and their splitting changes with temperature, it would provide direct evidence for the lifting of degeneracy due to octahedral tilts and rotations.

\indent In Fig.~\ref{T_dep}, we revisit the VBM at R and probe the R-M-R direction at 29 eV. The photon energy 29 eV is equivalent to the 54 eV used in Fig.~\ref{orbital}~{\bf b}, as both probe the R point \cite{Sajedi22}. Interestingly, we can probe  two nonequivalent M points, M and \Mp\ \oliv{(as denoted in Fig.~\ref{orth}~{\bf c}).}
\Mp\ is energetically separated from the VBM at R and appears in both Fig.~\ref{T_dep} and Fig.~\ref{orbital}~{\bf a,b}. \oliv{In Fig.~\ref{T_dep} it arises} from an umklapp from \Mp\ to R (vertical blue dashed arrow in Fig.~\ref{orth}~{\bf c}). With increasing temperature, the energy position of \Mp\ relative to the VBM shifts significantly, from 680~meV at 38~K to 400~meV at 300~K.   

In the panel centers of Fig.~\ref{T_dep}, we probe \oliv{the MX} band near M at approximately 0.2~MX, which requires a correction of about 150~meV 
\oliv{to determine the energy at M} (see Suppl. Note 6). Fig.~\ref{T_dep}~{\bf d} plots the temperature dependence of \Mp\ and M. We find that the energy difference between M and \Mp\ decreases with increasing temperature, confirming that the degeneracy is lifted at low temperatures. 
However, the fact that M and \Mp\ become degenerate as early as $\sim$200~K remains unexplained.
\oli{We include the data for \Mp\ from Fig.~\ref{orbital} measured under a different geometry at 39~eV. They confirm the much larger temperature  shift of \Mp\ relative to M but do not clarify at which temperature they become degenerate.}
We speculate that this could be influenced by surface effects.


In summary, we investigated the temperature evolution of  \CsPbBr3 using ARPES and XRD. Despite its orthorhombic structure below 360~K,  ARPES previously detected only cubic bands at room temperature. We reveal weak orthorhombic features   at 300~K, which become very pronounced at low temperatures, including a backfolded  VBM at $\Gamma$ and M. By tracking energy shifts with ARPES and performing temperature-dependent XRD, we establish a simulation model and compare ARPES data with DFT calculations based on XRD parameters.

Upon cooling, we observe a shift of \Mp\ that narrows the band along $\Gamma$\Mp\  (Fig.~\ref{orbital}) and widens it  along R\Mp\ (Fig.~\ref{T_dep}), a signature of the  second-order JT effect confirmed by our XRD-based DFT calculations. Additionally, the lifting of degeneracy at $\Gamma$ and between M and \Mp\ further supports this interpretation. To our knowledge, such shifts have not been reported in any halide or oxide perovskite. The fact that there is good agreement between our ARPES data and DFT calculations also supports a second-order Jahn–Teller effect. While we do not entirely exclude the possibility of a first-order effect, such behavior in perovskites has so far been linked to the strong electron correlation of transition-metal d-states, which typically defy a description by standard DFT.

\rader{We directly relate electronic structure changes in CsPbBr$_3$ to octahedral tilts and rotations, reconciling conflicting predictions --- from a 0.5 eV shift at M to claims of negligible structural impact. Our results reveal strong coupling between electronic states and lattice distortions, offering fundamental insights into electron–phonon interactions with implications for polaronic transport, dynamic disorder, defect tolerance, bandgaps, and exciton trapping.}
\rader{The observed band structure renormalization underscores the role of lattice dynamics in shaping the optoelectronic properties of halide perovskites. Temperature-dependent mobility could, in principle, distinguish between small and large polarons: the former show hopping transport and increasing mobility with temperature, while the latter imply enhanced effective mass. Although DFT predicts small polarons \cite{Neukirch16}, experimental trends favor large polarons \cite{Herz17,Bonn17}, yet our ARPES data revealed no mass enhancement at the VBM relative to state-of-the-art $GW$ theory \cite{Sajedi22}. A possible resolution may lie in recently predicted topologically nontrivial polarons \cite{Lafuente24}, which exhibit energy and length scales intermediate between small and large polarons. Our present study contributes to resolving the nature of the unconventional electron–lattice coupling at the heart of these predictions.}


 {\it Acknowledgments.} S.A.K. and J.M. thank the QM4ST project financed by the Ministry of Education of the Czech Republic grant no. CZ.02.01.01/00/22\_008/0004572, co-funded by the European Regional Development Fund.


\phantom{xxxx}

\bibliography{Jan_Teller}

  \end{document}